\documentclass[prd,nofootinbib,showpacs,preprint]{revtex4}
\usepackage{graphicx}
\usepackage{epsfig}
\usepackage{amsmath}
\usepackage{amsfonts}
\usepackage{amssymb}
\usepackage{url}
\usepackage{hyperref}
\usepackage{subfigure}
\usepackage{cancel}
\usepackage{eso-pic}
\usepackage[abs]{overpic}
\usepackage{xcolor,varwidth}
\newcommand{\beqa}{\begin{eqnarray}}
\newcommand{\eeqa}{\end{eqnarray}}
\newcommand{\beq}{\begin{equation}}
\newcommand{\eeq}{\end{equation}}

\newcommand{\bmt}{\begin{pmatrix}}
\newcommand{\emt}{\end{pmatrix}}
\usepackage[toc,page]{appendix}
\usepackage{comment}
\newcommand{\be}{\begin{equation}}
\newcommand{\ee}{\end{equation}}
\newcommand{\bea}{\begin{eqnarray}}
\newcommand{\eea}{\end{eqnarray}}

\usepackage{xspace}
\newcommand{\numu}{$\nu_\mu$\xspace}
\begin{document}
\title{Status of a Deep Learning Based Measurement of the Inclusive Muon Neutrino Charged-current Cross Section in the NOvA Near Detector}
\author{Biswaranjan Behera}
\email{bbehera@fnal.gov}
\affiliation{\,Department of Physics, IIT Hyderabad,
              Telangana - 502285, India \\
Fermilab, USA\\On behalf of the NOvA Collaboration\\ \\
              Talk presented at the APS Division of Particles and Fields Meeting (DPF 2017), July 31-August 4, 2017, Fermilab. C170731}

%%%%%%%%%%%%%%%%%%%%%%%%%%%%%%%%%%%%%%%%%%%%%%%%%%%%%%%%%%%%
\begin{abstract}
%%%%%%%%%%%%%%%%%%%%%%%%%%%%%%%%%%%%%%%%%%%%%%%%%%%%%%%%%
NOvA is a long-baseline  neutrino oscillation experiment. It uses the NuMI beam from Fermilab and two sampling calorimeter detectors placed off-axis from the beam. The 293 ton Near Detector measures the unoscillated neutrino energy spectrum, which can be used to predict the neutrino energy spectrum observed at the 14 kton Far Detector. The Near Detector also provides an excellent opportunity to measure neutrino interaction cross sections with high statistics, which will benefit current and future long-baseline neutrino oscillation experiments. This analysis implements new algorithms to identify $\nu_{\mu}$ charge-current events by using visual deep learning tools such as convolutional neural networks. We present the status of a measurement of the inclusive $\nu_{\mu}$ CC cross section in the NOvA Near Detector.    

\end{abstract}
\pacs{13.15.+g}
\maketitle
%%%%%%%%%%%%%%%%%%%%%%%%%%%%%%%%%%%%%%%%%%%%%%%%%%%%%%%%%%%%%%%%%%%%%%%%%%%%%%%%
\section{Introduction}
%%%%%%%%%%%%%%%%%%%%%%%%%%%%%%%%%%%%%%%%%%%%%%%%%%%%%%%%%%%

\subsection{Motivation}
The current and future generation of long-baseline neutrino oscillation experiments aim for the determination of the neutrino mass ordering (inverted vs. normal mass hierarchy);  precise measurement of $\theta_{23}$ (if it does not satisfy maximal, a determination of the octant it belongs to: $\theta_{23} < \frac{\pi}{4}  $ vs. $\theta_{23}  > \frac{\pi}{4}$); establishment of whether nature violates CP in the lepton sector. To achieve these goals we need a precision oscillation experiment. Understanding of the incoming neutrino energy, detector efficiency and modeling of neutrino-nucleus scattering is critical for such precise oscillation measurements. 

\begin{figure}[!htbp]
    \centering
        \includegraphics[width=0.9\textwidth]{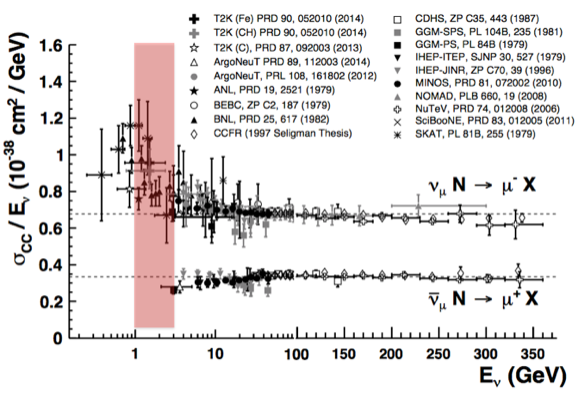}
    \caption{Neutrino charged-current inclusive cross section measurements as a function of neutrino energy, pink band represents NOvA's region of interest   \cite{pdgInclusiveXSec}.
   }
    \label{fig:pdgInclusiveXsec}
  \end{figure}

Charged-current (CC) interactions, with a charged-lepton in the final state, are used to measure 3-flavor oscillations probabilities. The final state lepton identifies the neutrino flavor and the energy of the neutrino can be reconstructed. However, the initially produced hadrons interact with dense nuclear matter within the nucleus, the final state interactions (FSI) can change the energy, angle and even the charge state of the originally produced hadrons (for example charged pion goes in to neutral pion and some pions get absorbed within the nucleus and do not emerge in the detector). Therefore, predicting the kinematic distributions of the final state leptons is critical. Such predictions rely on a-priori knowledge of neutrino-nucleus cross sections.  As shown in Fig.~\ref{fig:pdgInclusiveXsec}, neutrino-nucleus charged-current inclusive cross sections in the energies relevant for NOvA have 10-20\% uncertainties.

The NOvA Far Detector (FD) is 810 km from the NuMI production target and positioned 14 mrad off-axis from the NuMI beam, resulting in a narrow-band neutrino flux peaked around 2 GeV \cite{tdr}.  The NOvA Near Detector (ND) is located approximately 1 km from the NuMI production target, off-axis such that the peak of the neutrino flux matches that of the far detector. The high rate of neutrino interactions in the ND provides an opportunity for a rich program of neutrino-nucleus cross section measurements. Both detectors are functionally identical, segmented, tracking calorimeters. The basic unit of the NOvA detectors is a  highly reflective white polyvinyl chloride (PVC) cell of cross sectional size 3.9 cm by 6.6 cm and 4.2 m long in the near detector filled with liquid-scintillator. Each cell corresponds to one unique fiber and one unique pixel, 32 cells make up one module. A series of modules glued together into one large flat piece named as plane (1 plane = 12 modules in FD and 3 modules in ND fiducial volume). The detectors are designed to provide sufficient sampling of hadronic and electromagnetic showers to allow efficient separation of the charged current (CC) interaction signals from the neutral current (NC) interactions.

\subsection{Analysis Overview}
We aim to measure the muon-neutrino charged-current inclusive cross section in the NOvA near detector in bins of true neutrino energy:
\begin{equation}
\sigma(E_i) = \frac{\sum_j U_{ij}\left(N_\mathrm{sel}(E_j) - N_\mathrm{bkg}(E_j)\right)}{\epsilon(E_i)N_\mathrm{target}\Phi(E_i)}
\label{eq:xsec}
\end{equation}
as well as flux-integrated double-differential cross section with respect to the final-state muon's true kinetic energy and true angle:
\begin{equation}
\left(\frac{d^2\sigma}{d\cos\theta_{\mu} dT_{\mu}}\right)_i = \frac{\sum_{j} U_{ij}(N^\mathrm{sel}(\cos\theta_{\mu}, T_{\mu})_j - N^\mathrm{bkg}(\cos\theta_{\mu}, T_{\mu})_j)}{\epsilon(\cos\theta_{\mu}, T_{\mu})_i  (\Delta\cos\theta_\mu)_i (\Delta T_\mu)_i N_\mathrm{target}\Phi}
\label{eq:xsec_doublediff}
\end{equation}
Here, $N^\mathrm{sel}$ is the number of selected events, $N^\mathrm{bkg}$ is the estimated number of background events, and $U$ is the unfolding matrix that corrects the reconstructed quantities for detector resolution, acceptance and efficiency, $\Phi$ is the neutrino flux, $\epsilon$ is signal selection efficiency and $N_\mathrm{target}$ is the number of target nucleons in the fiducial volume. This measurement is of interest to the general neutrino community, and the measured kinematic distributions of the final-state muon can be used to improve the simulated neutrino interactions in the NOvA detectors.  The inclusive cross section measurement may serve as a basis for future muon-neutrino semi-inclusive cross section measurements, in particular where the dominant flux systematic uncertainty can be mitigated via the ratio of semi-inclusive to inclusive measurements.  
The signal in this measurement is defined as all \numu CC interactions where the true neutrino energy is between 0.75 and 4 GeV and the true interaction vertex is in a well-defined fiducial volume. All interactions other than \numu CC or those \numu CC interactions that occur outside of the fiducial volume are regarded as background.
 
\section{Event Selection \label{sec:event_selection}}
This analysis requires selection of $\nu_\mu$ CC signal events. A pre-selection of the data requires at least one reconstructed track (trajectory of particle through detector), and filters out events with less than 20 hits (activity on a particular cell). The starting point of the reconstructed muon track is assumed to be the vertex of the $\nu_\mu$ CC interaction and must fall inside our fiducial volume. In addition, we remove events where the muon track is not fully contained in the detector. Our first goal to select a $\nu_\mu$ CC event is to identify a muon in the event (individual fundamental interactions). To identify a muon track in the reconstructed $\nu_\mu$ CC event we use a multivariate analysis implementing a k-Nearest Neighbor algorithm. The resulting muon identification (Muon ID) is based on the measured dE/dx, amount of multiple scattering along the track, total track length, and the fraction of track planes that have overlapping hadronic activity. An event is selected or rejected as a charged-current interaction based on the highest Muon ID value of all reconstructed tracks in the event, and the track associated with that Muon ID is identified as the muon. We determine the Muon ID selection cut that optimizes the statistical figure of merit $s/\sqrt{s+b}$,  where $s$ is the number of signal events passing the cut and $b$ is the number of background events passing the cut. Backgrounds include  NC, beam $\nu_{e}$ CC and $\bar{\nu}_\mu$ CC interactions. The optimal Muon ID cut was found to be 0.29.

\begin{figure}[!htbp]
    \centering
    \begin{overpic}[width=0.5\textwidth]{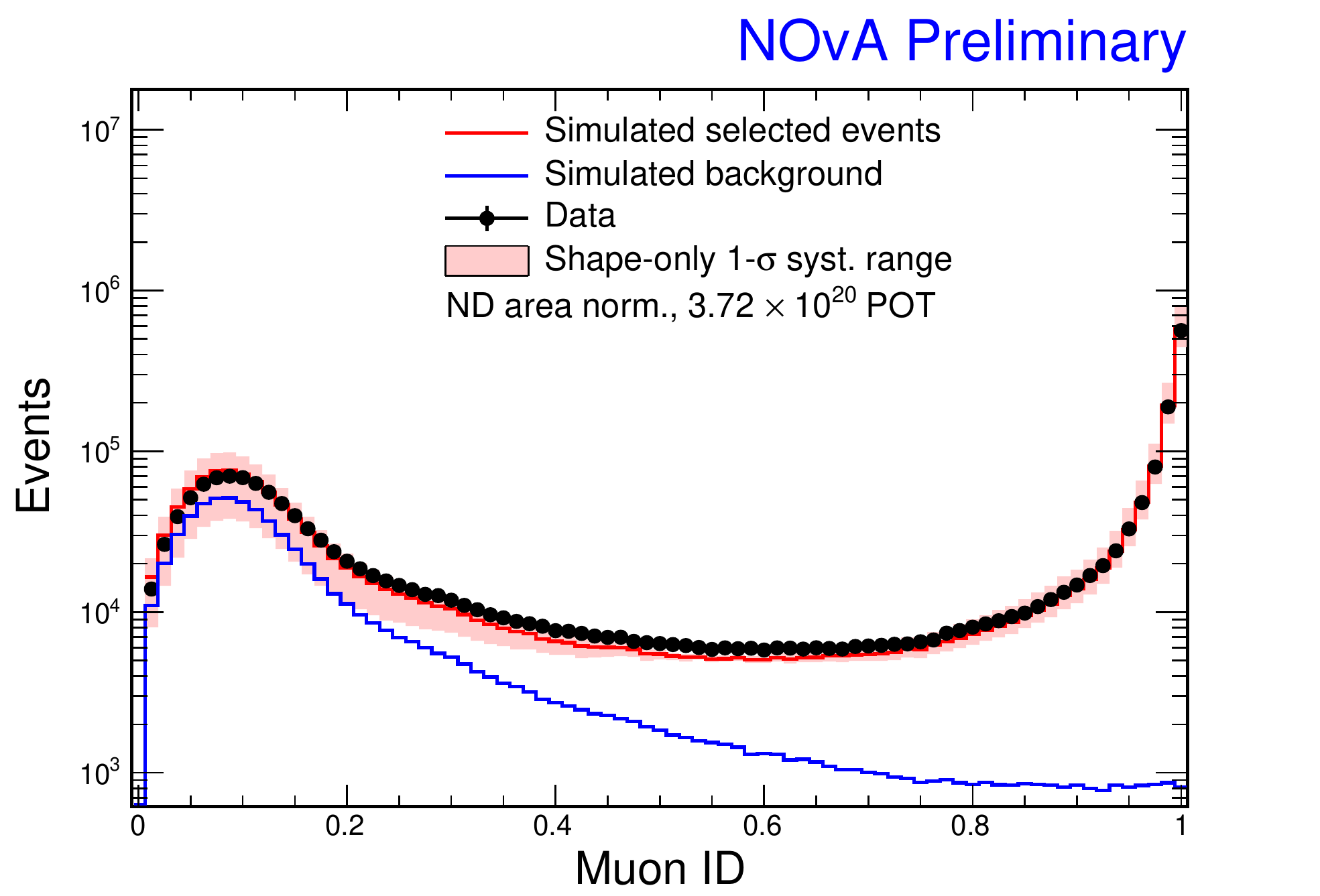}
\linethickness{0.5pt}
\put(78.,75){\color{magenta}\vector(1,0){10}}
\put(78.,15.3){\color{magenta}\line(0,1){60}}
\end{overpic}
    \caption{ Selecting events $>$ 0.29 are retained as $\nu_\mu$ CC events (Muon ID distribution).}
    %\cite{REMID}}
   \label{fig:REMID}
  \end{figure}    
%%%%%%%%%%%%%%%%%%%%%%%%%%%%%%%%%%%%%%a%%%%%%%%%%%%%%%%%%%%%%%%%

\section{Deep Learning Based Event Selection}
Convolutional neural networks (CNNs) have been widely applied in the machine learning community to solve complex problems in image recognition and analysis. CNN technology can also be applied to the problem of identifying particle interactions in the NOvA sampling calorimeters commonly used in high energy physics \cite{cvn}. The NOvA CNN algorithm, CVN (Convolutional Visual Network) identifies neutrino interactions based on their topology without the need for detailed reconstruction currently used by the NOvA experiment. The network is trained on two dimensional views of the event's calibrated hits and the output from each view is then combined in the final layers of the network. The use of CVN to select $\nu_{e}$ interactions for the appearance measurement in the NOvA FD resulted in an effective increase in exposure of 30\% compared to traditional event identification methods \cite{cvn}. 

\begin{figure}[!htbp]
    \centering
        \includegraphics[width=0.45\textwidth]{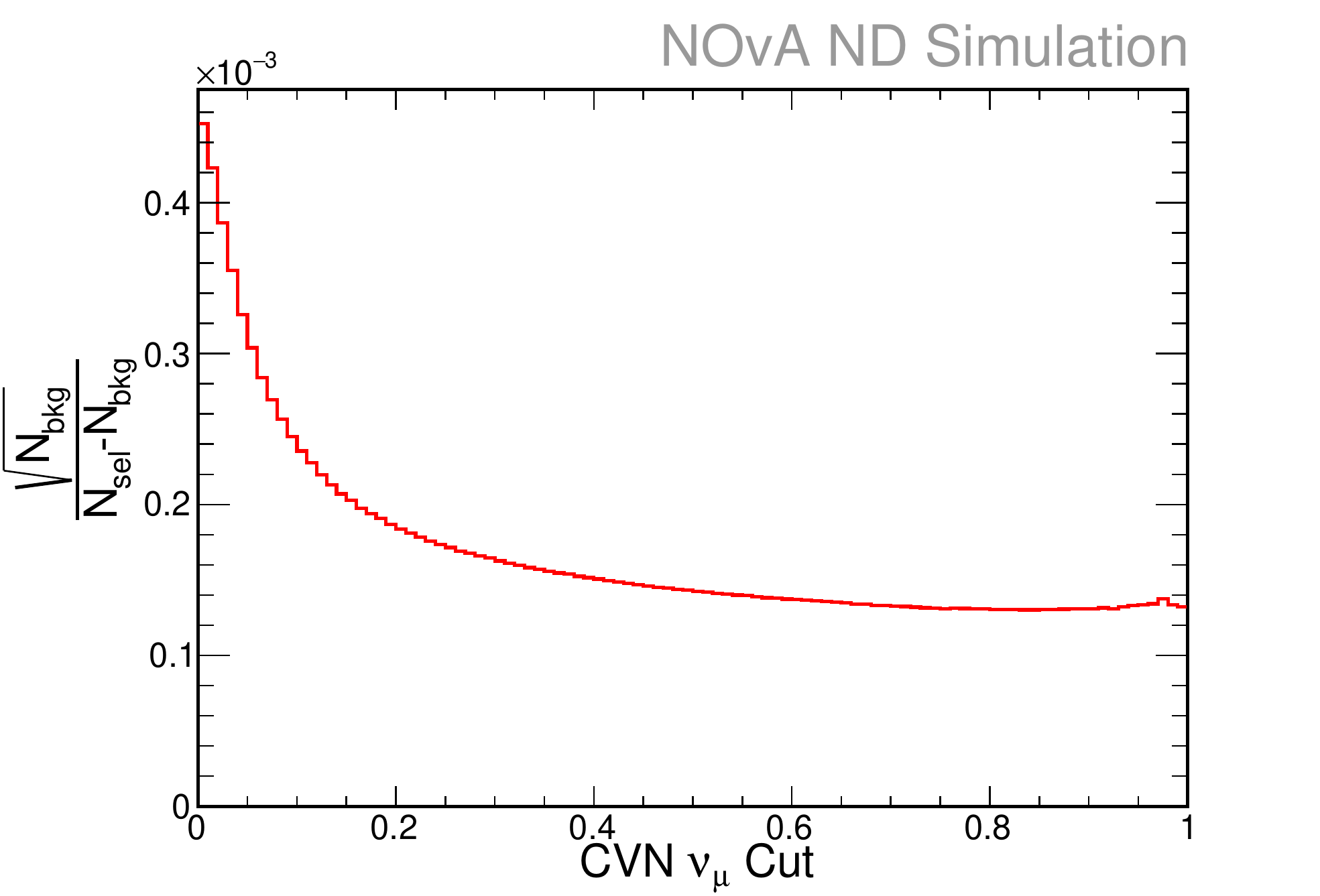}
    \includegraphics[width=0.45\textwidth]{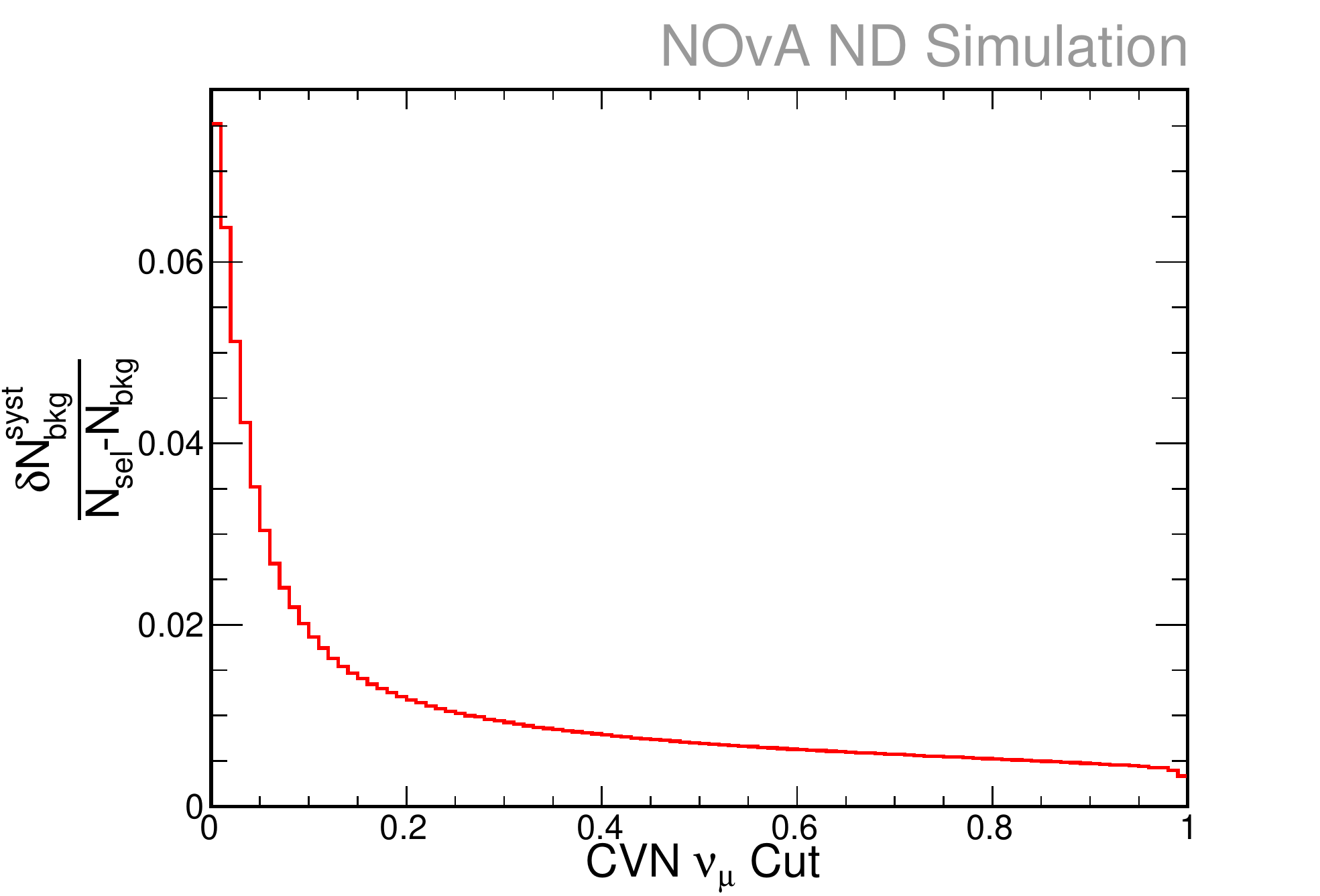}\\
    \includegraphics[width=0.45\textwidth]{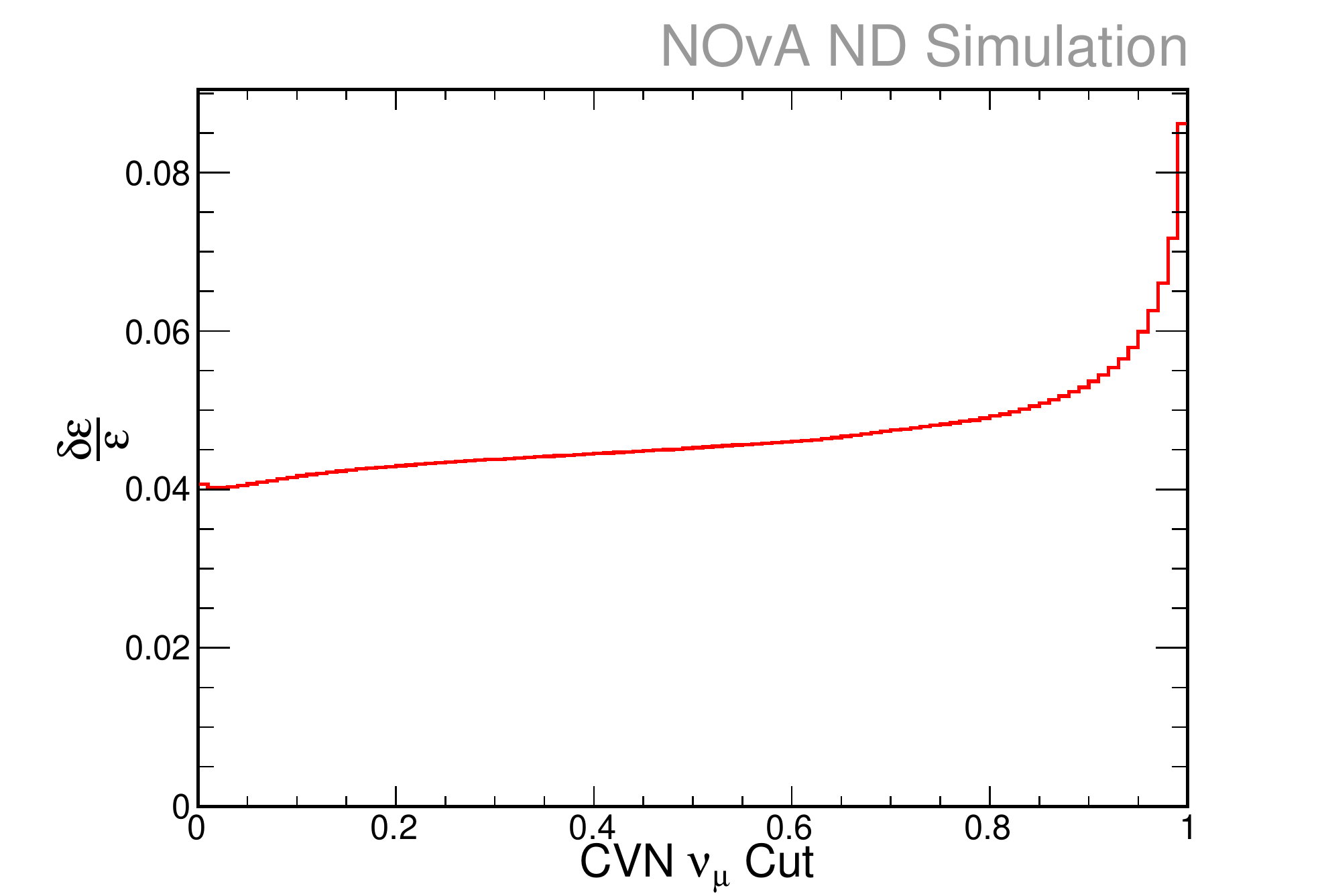}
    \begin{overpic}[width=0.45\textwidth]{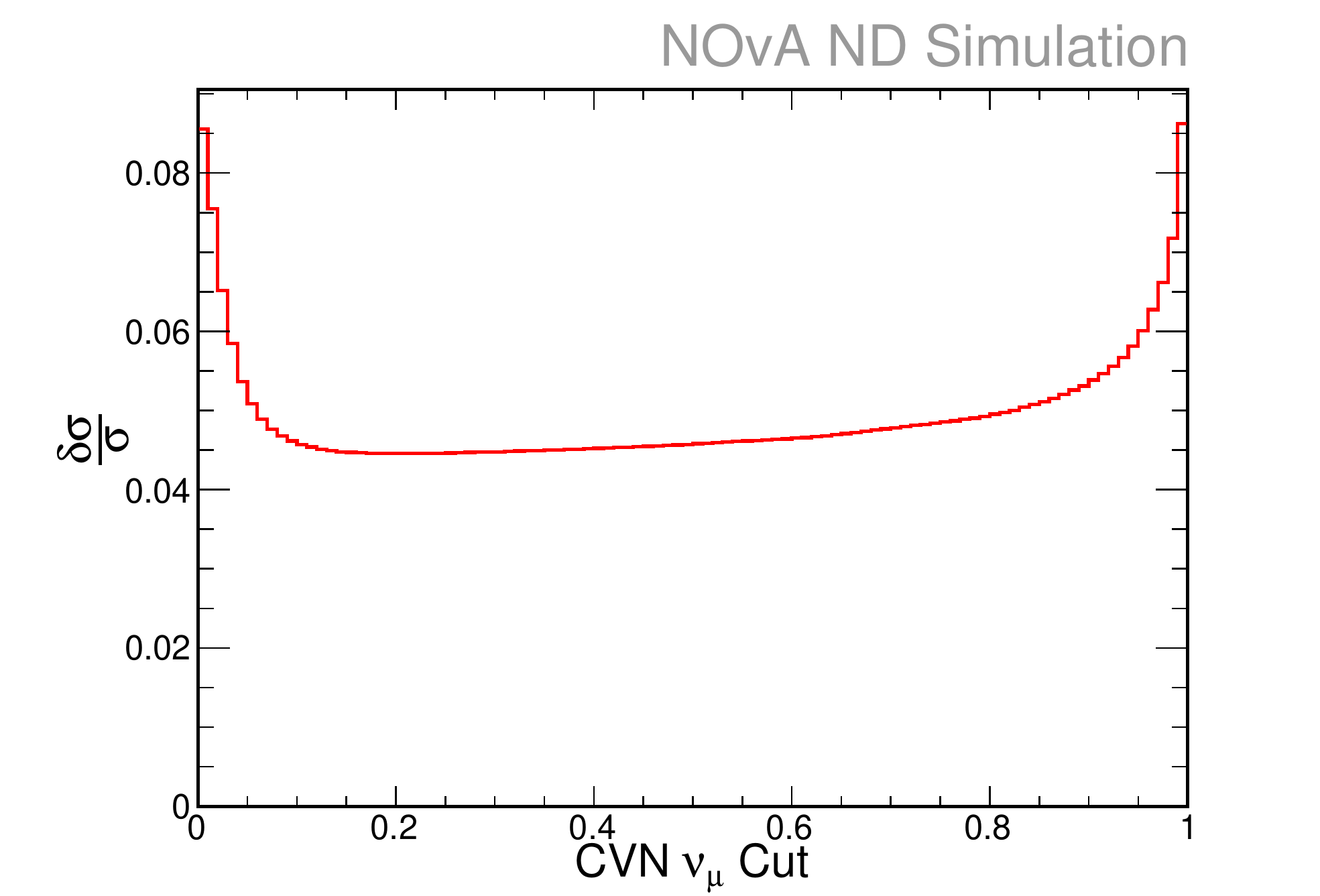}
\linethickness{0.5pt}
\put(64.,95){\color{cyan}\vector(1,0){20}}
\put(64.,15.3){\color{cyan}\line(0,1){80}}
\end{overpic}
    \caption{Fractional statistical uncertainty on background events (top left). fractional systematic uncertainty on background events (top right). Fractional uncertainty on efficiency (bottom left). fractional uncertainty on cross-section (bottom right).}%\cite{pdgInclusiveXSec}}
    \label{fig:sigma_unc}
  \end{figure}

We have developed a technique to select $\nu_{\mu}$ CC interactions in the NOvA ND using CVN optimized to minimize of the fractional systematic uncertainties on the inclusive $\nu_{\mu}$ CC cross section from the background estimation and efficiency (Fig \ref{fig:sigma_unc}): 
\begin{equation}
\frac{\delta\sigma}{\sigma} =\sqrt{\frac{\text N_{\text {bkg}} + (\delta \text N^\text {syst}_{\text {bkg}})^2}{(\text N_{\text {sel}} - \text N_{\text {bkg}})^2} + \left(\frac{\delta\epsilon}{\epsilon}\right)^2}
\label{eq:sigma_uncertaintity}
\end{equation}

\begin{figure}[!htbp]
    \centering
        \includegraphics[width=0.5\textwidth]{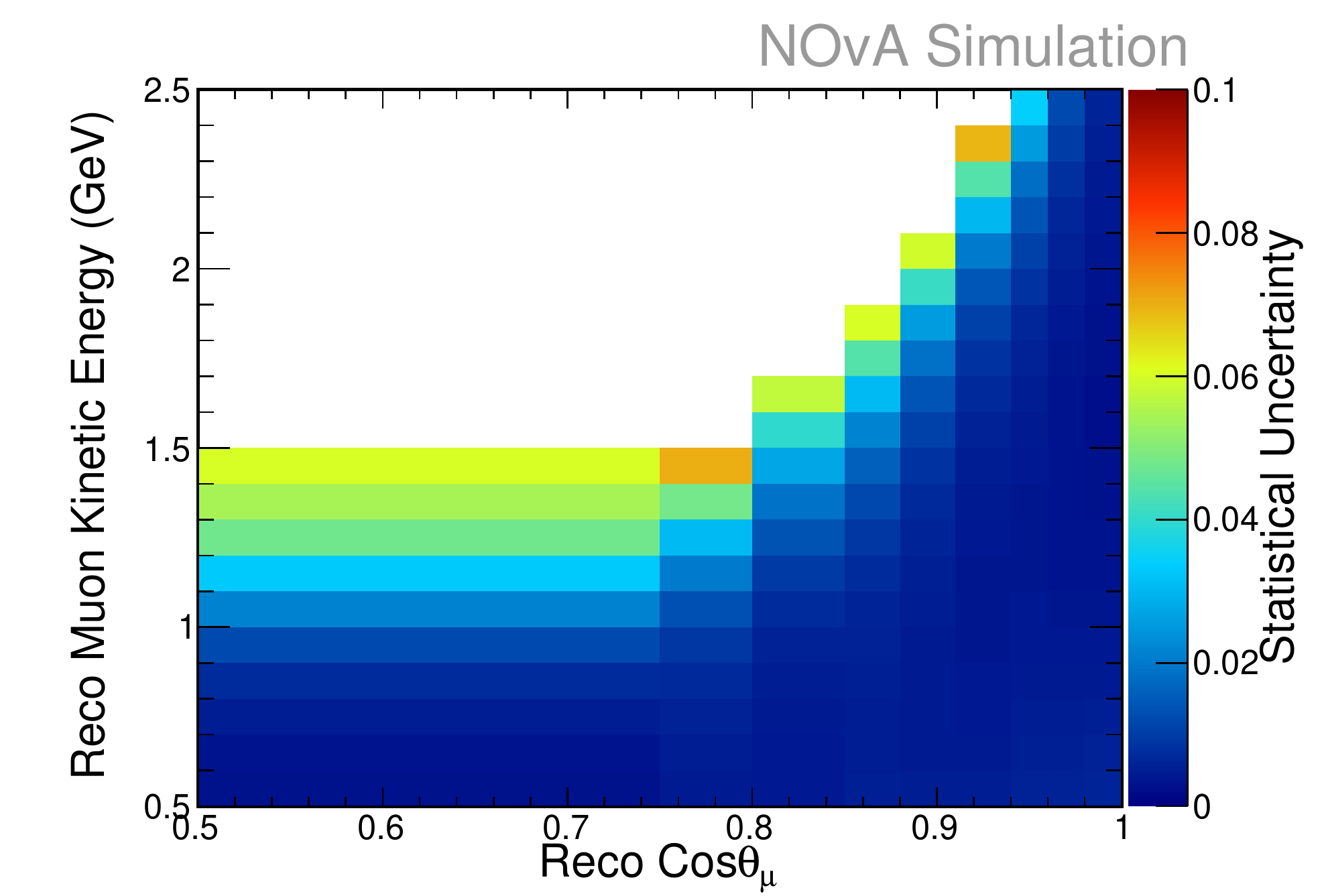}
    \caption{Statistical uncertainty for kinematic phase space }%\cite{pdgInclusiveXSec}}
    \label{fig:stat_unc}
  \end{figure}

The systematic uncertainty is estimated by comparison of modified Monte Carlo (MC) with our nominal Monte Carlo. The dominant pieces of the systematic uncertainty for this analysis are: the flux uncertainty, determined from comparisons to external hadron production data \cite{ppfx}, Cross-section and Final State Interaction (FSI) uncertainty are determined using GENIE re-weighting scheme\cite{genie}, and the hadronic energy uncertainty. The statistical uncertainty is very small, less than 1\%, for most of the bins is shown in Fig. \ref{fig:stat_unc}.

\section{Summary}
In summary, we will use the NOvA ND to measure the $\nu_\mu$ CC cross-section in addition to neutrino oscillation studies. However, the flux uncertainty is dominant ($\sim$ 8-10\%), compared to non-flux uncertainties ($\sim $5\%). We have developed a CVN based approach to select muon neutrino charge current events. Optimization of muon paricle identification and estimation of the measured neutrino energy is in progress.
 
\section{ACKNOWLEDGMENTS}
This work was supported by the US Department of Energy; the US National Science Foundation; the Department of Science and Technology, India; the European Research Council; the MSMT CR, Czech Republic; the RAS, RMES, and RFBR, Russia; CNPq and FAPEG, Brazil; and the State and University of Minnesota. We are grateful for the contributions of the staffs of the University of Minnesota module assembly facility and NOvA FD Laboratory, Argonne National Laboratory, and Fermilab. Fermilab is operated by Fermi Research Alliance, LLC under
Contract No. DE-AC02-07CH11359 with the U.S. Department of Energy, Office of Science, Office of High Energy Physics. 
%%%%%%%%%%%%%%%%%%%%%%%%%%%%%%%%%%%%%%%%%%%%%%%%%%%%%%%

%%%%%%%%%%%%%%%%%%%%%%%%%%%%%%%%%%%%%%%%%%%%%%%%%%%%%%%%%%%%%%

%%%%%%%%%%%%%%%%%%%%%%%%%%%%%%%%%%%%%%%%%%%%%%%%%%%%%%%%%
%\begin{figure}[h]
%\centering
%\includegraphics[scale=0.55]{buenu.pdf}
%\quad
%\includegraphics[scale=0.55]{buenu.pdf}
%\quad
%\includegraphics[scale=0.55]{buenu.pdf}
%\caption{Constraint on the  real and imaginary part of the  $V_1$ Wilson coefficient obtained  from  $B_u^+ \to e^+ \nu_e$ (left panel), $B_u^+ \to \mu^+ \nu_\mu$ (right panel) and $B_u^+ \to \tau^+ \nu_\tau$ (bottom panel) processes. }
%\end{figure}
%%%%%%%%%%%%%%%%%%%%%%%%%%%%%%%%%%%%%%%%%%%%%%%%%%%%%%%%%

%%%%%%%%%%%%%%%%%%%%%%%%%%%%%%%%%%%%%%%%%%%%%%%%%%%%%%%%%

%%%%%%%%%%%%%%%%%%%%%%%%%%%%%%%%%%%%%%%%%%%%%%%%%%%%%%%

%%%%%%%%%%%%%%%%%%%%%%%%%%%%%%%%%%%%%%%%%%%%%%%%%%%%%%%%%%%

%%%%%%%%%%%%%%%%%%%%%%%%%%%%%%%%%%%%%%%%%%%%%%%%%%%%%%%%%%%

\end{document}